\documentclass[reprint,groupedaddress,showpacs,preprintnumbers,
amsmath,amssymb,aps,prb]{revtex4-1}
\usepackage{graphicx}
\usepackage{dcolumn}
\usepackage{bm}
\usepackage{mathrsfs}
\usepackage{color}
\begin{document}
\newcommand{\equref}[1]{Eq.\ (\ref{#1})}
\newcommand{\figref}[1]{FIG.\ \ref{#1}.}
\newcommand{\tabref}[1]{TABLE\ \ref{#1}.}
\title{Josephson effect in a multi-orbital model for Sr$_{2}$RuO$_{4}$}


\author{Kohei Kawai}
\affiliation{Department of Applied Physics, Nagoya University, Nagoya 464-8603, Japan}

\author{Keiji Yada}
\affiliation{Department of Applied Physics, Nagoya University, Nagoya 464-8603, Japan, Moscow Institute of Physics and Technology, Dolgoprudny, Moscow 141700, Russia}

\author{Yasuhiro Asano}
\affiliation{Department of Applied Physics, Hokkaido University, Sapporo 060-8628, Japan, Moscow Institute of Physics and Technology, Dolgoprudny, Moscow 141700, Russia}

\author{Alexander A. Golubov}
\affiliation{Faculty of Science and Technology and MESA+ Institute of Nanotechnology,
University of Twente, 7500 AE, Enschede, The Netherlands, Moscow Institute of Physics and Technology, Dolgoprudny, Moscow 141700, Russia}

\author{Satoshi Kashiwaya}
\affiliation{National Institute of Advanced Industrial Science and Technology (AIST), Tsukuba 305-8568, Japan, Moscow Institute of Physics and Technology, Dolgoprudny, Moscow 141700, Russia}

\author{Yukio Tanaka}
\affiliation{Department of Applied Physics, Nagoya University, Nagoya 464-8603, Japan, Moscow Institute of Physics and Technology, Dolgoprudny, Moscow 141700, Russia}


\date{\today}

\begin{abstract}
We study Josephson currents between $s$-wave/spin-triplet superconductor junctions by
taking into account details of the band structures in Sr$_{2}$RuO$_{4}$, such as three conduction bands and
spin-orbit interactions in the bulk and at the interface. We assume five superconducting order parameters 
in Sr$_{2}$RuO$_{4}$: a chiral $p$-wave symmetry and four helical $p$-wave symmetries. 
We calculate the current-phase relationship $I(\varphi)$ in these junctions, 
where $\varphi$ is the macroscopic phase difference between the two superconductors. The results for a chiral $p$-wave pairing symmetry show that a
 $\cos(\varphi)$ term appears in the current-phase relation because of time-reversal symmetry (TRS) breaking.
On the other hand, this $\cos(\varphi)$ term is absent in the helical pairing states that preserve TRS. 
We also study the dependence of the maximum Josephson current $I_c$ on an external magnetic flux $\Phi$ in a corner junction.
The calculated $I_c(\Phi)$ obeys
$I_{c}(\Phi) \neq I_{c}(-\Phi)$ in a chiral state and
$I_{c}(\Phi)=I_{c}(-\Phi)$ in a helical state.
We calculate $I_c(\Phi)$ in a corner SQUID and a symmetric SQUID geometry.
In the latter geometry, 
$I_{c}(\Phi)=I_{c}(-\Phi)$ is satisfied for all the pairing states 
and it is impossible to distinguish a chiral state from a helical one. 
On the other hand, a corner SQUID always 
gives $I_{c}(\Phi) \neq I_{c}(-\Phi)$ and
$I_{c}(\Phi)=I_{c}(-\Phi)$ for a chiral and a helical state, respectively.
Experimental tests of these relations in corner junctions and SQUIDs may serve as a tool for unambiguously determining the pairing symmetry 
in Sr$_{2}$RuO$_{4}$.
\end{abstract}

\maketitle


\section{Introduction}\label{sec1}
Strontium ruthenate (Sr$_{2}$RuO$_{4}$, or SRO) has attracted much interest for its unconventional superconductivity below the critical temperature $T_{c} \sim 1.5$ K \cite{Maeno94}.
The constancy of the Knight shift across $T_c$ is strongly indicative of 
spin-triplet pairing order \cite{Ishida,Maeno98,Murakawa,Maeno2,Maeno2012}.
Many theoretical studies have examined the microscopic 
mechanism of spin-triplet pairings in this material 
\cite{Rice,Miyake,Ogata,Nomura00,Nomura02a,Nomura02b,Arita,Nomura05,Nomura08,Yanase,Raghu,Kohmoto,Kuroki,Takimoto,Onari}. 
Exotic phenomena 
specific to spin-triplet superconductors \cite{Proximityp,Proximityp2,Proximityp3,Meissner3,odd1} are therefore naturally expected 
in SRO. Although several studies have focused on the superconducting order parameter, 
the symmetry of a Cooper pair is not yet fully understood.
Five spin-triplet pairing states are compatible with the tetragonal crystal structure of SRO \cite{Maeno2}.
One of these is a spin-triplet chiral $p$-wave state
(denoted the $E_{u}$ state in the Mulliken notation) where the $d$-vector is parallel to $c$-axis of the crystal.
The other four candidates are called spin-triplet helical states (denoted $A_{1u}$, $A_{2u}$, $B_{1u}$, and $B_{2u}$ in the Mulliken notation), 
where the $d$-vectors lie in the $ab$-plane of the crystal.\par
According to the recently proposed topological classification\cite{Schnyder08,qi11,tanaka12,alicea12},
all of the proposed superconducting states are topologically nontrivial.
Consequently, topologically protected Andreev bound states are expected at an SRO surface \cite{Kashiwaya11}.
Some experimental results are consistent with the proposed pair potential.
It has been suggested that the maximum Josephson current in Au$_{0.5}$In$_{0.5}$-SRO superconducting quantum interference devices (SQUID) displays an odd-parity pairing state \cite{Nelson}. \par
Tunneling spectroscopy experiments also suggest the formation of
a dispersive surface Andreev bound state (SABS) at the in-plane edges of SRO \cite{Laube,Kashiwaya11,Liu}.
The dispersive SABSs \cite{Furusaki2001,Sengupta} are distinguishable from the dispersionless SABS in a $d$-wave superconductor.
The former generates a broad zero-bias conductance peak
(ZBCP) \cite{YTK97,YTK98,Honerkamp}, whereas the latter forms a sharp ZBCP \cite{TK95,Kashiwaya00,Hu}.
Because SRO is a multi-band superconductor, 
the numerically determined 
energy dispersion of an SABS in a multi-band model 
is more complicated than that in a single-orbital model  
\cite{Imai2012,Imai2013}. 
Yada, \textit{et al.} 
successfully explained the variety of conductance spectra observed in experiments 
\cite{Kashiwaya11} 
in terms of the three-band degrees of freedom \cite{Yada}. 
Several Josephson-junction experiments suggested the presence of domain structures, detected from an anomalous current-switching behavior \cite{Kidwingira,Kambara1,Kambara2,Anwar,Saito}.
These experimental findings are consistent with the existence of both chiral and helical $p$-wave pairing symmetries in SRO.

A chiral state is qualitatively different from the four helical states because 
it breaks the time-reversal symmetry (TRS), whereas the helical states preserve TRS \cite{Qi09}.
Although the presence or absence of TRS in SRO is an important issue,
experimental results remain controversial.
TRS breaking can be verified by observing a spontaneous magnetic field or a spontaneous 
edge current. Theoretical studies have shown 
that the amplitude of the spontaneous magnetization is detectable 
experimentally \cite{Matsumoto99}
and that the edge current is robust 
with respect to surface roughness \cite{Suzuki}. 
%
Measurements of muon spin resonance and of the Kerr effect have detected the presence of an internal magnetic field \cite{Luke,Xia}, which in turn suggests a chiral $p$-wave symmetry.
On the other hand, scanning SQUID experiments have not shown any signs of a spontaneous magnetic field \cite{Kirtley,Hicks},
which suggests a helical $p$-wave symmetry.
Several theoretical proposals have been put forward to explain the absence of the edge current in SRO \cite{Kallin1,Kallin2,Kallin3,Raghu,Tada}.
A resolution of this paradox requires an experimental test able to distinguish unambiguously between a chiral and 
a helical pairing symmetry.

In this paper, we present a theory of the Josephson effect between
a spin-singlet $s$-wave superconductor and a spin-triplet 
$p$-wave superconductor 
by taking into account the three bands of the SRO in addition to the spin-orbit interaction in the bulk and at the interface.
The importance of multi-orbital effects are apparent in various 
physical quantities \cite{Nomura04,Nomura08}. 
Since spin-orbit coupling influences the current-phase relation 
fundamentally, it is necessary that our theory consider a three-band model.  
We calculated the current-phase relation $I(\varphi)$ in Josephson junctions, 
where $\varphi$ is the macroscopic phase difference between the two superconductors. We found that $\cos(\varphi)$ appears in $I(\varphi)$ for chiral $p$-wave pairing, owing to TRS breaking, 
to ensure consistency with previous results \cite{asano03}. However, $\cos(\varphi)$ is absent for helical pairing, thus reflecting time-reversal invariance. In the case of helical pairing, $\sin(\varphi)$ appears only 
in a three-band model. 
We also studied the dependence of the maximum Josephson
current $I_{c}$ on an external magnetic flux $\Phi$ in two types of SQUID geometries: a corner SQUID and a symmetric SQUID.
In a corner Josephson junction and a corner SQUID,
we found $I_{c}(\Phi) \neq I_{c}(-\Phi)$ for a chiral state, whereas 
$I_{c}(\Phi)=I_{c}(-\Phi)$ holds true for a helical state.
We show that the three-band character affects the oscillation period of 
$I_{c}(\Phi)$. 
It is possible to determine the pairing symmetry unambiguously by testing these relations in SRO-based corner junctions and SQUIDs.
In a symmetric SQUID, the relation $I_{c}(\Phi)=I_{c}(-\Phi)$ is satisfied in both chiral and helical cases.

\section{Model and Formulations}\label{sec2}
This section introduces a model Hamiltonian for an SRO/normal metal (NM)/$s$-wave superconductor junction system.
First, we explain the Hamiltonian for bulk SRO, which consists of three terms $\mathcal{H}_{\rm kin}$, $\mathcal{H}_{\rm soi}$ and $\mathcal{H}_{\rm pair}$.
The first term $\mathcal{H}_{\rm kin}$ expresses the kinetic energy.
ARPES measurements and first-principles calculations have shown that SRO has three two-dimensional Fermi surfaces \cite{Oguchi,Singh,Cuoco,Haverkort}.
These Fermi surfaces were reproduced by considering three orbitals, $i.e.$, the $d_{xy}$, $d_{yz}$, and $d_{zx}$ orbitals, in SRO.
We can therefore consider a three-band two-dimensional Hamiltonian constructed using the tight-binding model:
\begin{equation}
\mathcal{H}_{\rm kin}=\sum_{{\bm k},\sigma}\hat c^\dag_{{\bm k}\sigma}
\begin{pmatrix}
\varepsilon_{yz}({\bm k})&g({\bm k})&0\\
g({\bm k})&\varepsilon_{zx}({\bm k})&0\\
0&0&\varepsilon_{xy}({\bm k})
\end{pmatrix}c_{{\bm k}\sigma},\label{hkin_equ}
\end{equation}
where ${\bm k}$ is a wavenumber, $\sigma$ is the spin, and $\hat c_{{\bm k}\sigma}=(c_{{\bm k},\sigma}^{yz}, c_{{\bm k},\sigma}^{zx}, c_{{\bm k},-\sigma}^{xy})^T$ is the annihilation operator.
The matrix components of \equref{hkin_equ} are given by
\begin{align}
\varepsilon_{xy}({\bm k})&=-2t_1(\cos k_x+\cos k_y)-4t_2\cos k_x\cos k_y-\mu_{xy},\\
\varepsilon_{yz}({\bm k})&=-2t_4\cos k_x-2t_3\cos k_y-\mu_{yz},\\
\varepsilon_{zx}({\bm k})&=-2t_3\cos k_x-2t_4\cos k_y-\mu_{zx},\\
g({\bm k})&=-4t_5\sin k_x\sin k_y,
\end{align}
where $t_1$, $t_2$, $t_3$, $t_4$, and $t_5$ are the hopping integrals up to next nearest-neighbor sites.
The second term $\mathcal{H}_{\rm soi}$ denotes the spin-orbit interaction in bulk SRO,
\begin{eqnarray}
\mathcal{H}_{\rm soi}&=\lambda\sum_{{\bm k},\sigma}
\hat c^\dag_{{\bm k}\sigma}
\begin{pmatrix}
0&is_\sigma&-s_\sigma\\
-is_\sigma&0&i\\
-s_\sigma&-i&0
\end{pmatrix}
\hat c_{{\bm k}\sigma},
\end{eqnarray}
where $s_\sigma=1$ ($s_\sigma=-1$) for $\sigma=\uparrow$ ($\sigma=\downarrow$).
This term mixes the spin and orbital degrees of freedom.
The third term $\mathcal{H}_{\rm pair}$ expresses the pair potential in SRO.
We chose spin-triplet chiral and helical $p$-wave pairings in the following analysis.
In the chiral $p$-wave case, we considered a pair potential which belongs
to the $E_u$ irreducible representation.
In the helical $p$-wave case, we considered two kinds of pair potentials
belonging to the $A_u$ and $B_u$ irreducible representations.
Using the orbital-dependent $d$ vector $d^{\ell}(\bm k)$,
the pair potential can be expressed as
\begin{eqnarray}
	\mathcal{H}_{\rm pair}=\sum_{\ell}{\hat c^{\ell\dagger}\left( \begin{array}{cc}
\hat0 & \hat\Delta^{\ell}(\bm k) \\
-\hat\Delta^{\ell\ast}(-\bm k) & \hat0 \\
\end{array} \right)\hat c^{\ell}},
\end{eqnarray}
with $\hat c^{\ell}=(c_{{\bm k},\uparrow}^{\ell}, c_{{\bm k},\downarrow}^{\ell}, c_{{-\bm k},\uparrow}^{\ell\dagger}, c_{{-\bm k},\downarrow}^{\ell\dagger})^T$ , and $\hat{\Delta}^{\ell}(\bm k)=i\bm d^{\ell}(\bm k)\cdot\bm\sigma\sigma_y$, 
where $\ell$ denotes the orbital index.
The five kinds of $d$ vectors are given by
\begin{align}
&\begin{cases}
\bm d_{Eu}^{yz}=\hat z\Delta_1(\delta\sin{k_x}+i\sin{k_y}),\\
\bm d_{Eu}^{zx}=\hat z\Delta_1(\sin{k_x}+i\delta\sin{k_y}),\\
\bm d_{Eu}^{xy}=\hat z\Delta_2(\sin{k_x}+i\sin{k_y}),
\end{cases}\\
&\begin{cases}
\bm d_{A1u}^{yz}=\hat x\delta\Delta_1\sin{k_x}+\hat y\Delta_1\sin{k_y},\\
\bm d_{A1u}^{zx}=\hat x\Delta_1\sin{k_x}+\hat y\delta\Delta_1\sin{k_y},\\
\bm d_{A1u}^{xy}=\hat x\Delta_2\sin{k_x}+\hat y\Delta_2\sin{k_y},
\end{cases}\\
&\begin{cases}
\bm d_{A2u}^{yz}=\hat x\Delta_1\sin{k_y}-\hat y\delta\Delta_1\sin{k_x},\\
\bm d_{A2u}^{zx}=\hat x\delta\Delta_1\sin{k_y}-\hat y\Delta_1\sin{k_x},\\
\bm d_{A2u}^{xy}=\hat x\Delta_2\sin{k_y}-\hat y\Delta_2\sin{k_x},
\end{cases}\\
&\begin{cases}
\bm d_{B1u}^{yz}=\hat x\Delta_1\sin{k_x}-\hat y\delta\Delta_1\sin{k_y},\\
\bm d_{B1u}^{zx}=\hat x\delta\Delta_1\sin{k_x}-\hat y\Delta_1\sin{k_y},\\
\bm d_{B1u}^{xy}=\hat x\Delta_2\sin{k_x}-\hat y\Delta_2\sin{k_y},
\end{cases}\\
&\begin{cases}
\bm d_{B2u}^{yz}=\hat x\delta\Delta_1\sin{k_y}+\hat y\Delta_1\sin{k_x},\\
\bm d_{B2u}^{zx}=\hat x\Delta_1\sin{k_y}+\hat y\delta\Delta_1\sin{k_x},\\
\bm d_{B2u}^{xy}=\hat x\Delta_2\sin{k_y}+\hat y\Delta_2\sin{k_x}.
\end{cases}
\end{align}
In these pair potentials, we only considered the intra-orbital pairing cases.
Furthermore, we introduced anisotropy in the pair potential in quasi-one-dimensional $d_{yz}$ and $d_{zx}$ orbitals by setting $\delta<1$.
In addition, the crystalline symmetry of SRO allows different magnitudes of the pair potential for the two-dimensional $d_{yz}$ orbital ($\Delta_1$) and the quasi-one-dimensional $d_{yz}$ and $d_{zx}$ orbitals ($\Delta_2$).

In the NM region between an SRO and an $s$-wave superconductor, we considered a single-orbital model given by
\begin{align}
\mathcal{H}_{\rm NM}=\sum_{{\bm k}\sigma} (\varepsilon_{\bm k}-\mu)c^\dag_{{\bm k}\sigma}c_{{\bm k}\sigma},\label{eq:nm}
\end{align}
where $c_{{\bm k}\sigma}$ is the annihilation operator for an electron in the NM.
The energy dispersion of the NM is given by 
\[
\varepsilon_{k}=-2t_{1}(\cos(k_x)+\cos(k_y))-4t_{2}\cos(k_{x})\cos(k_{y}) 
-\mu_{n} 
\]
where $t$ is the hopping integral between nearest-neighbor sites. 
We took into account the interface Rashba spin-orbit coupling 
in the NM layer next to the SRO, which is given by
\begin{align}
\mathcal{H}_{\rm RSOI}=\lambda_R\sin k_y\hat\sigma_z.
\end{align}
In the spin-singlet $s$-wave superconductor region, we considered the on-site pair potential as well as the kinetic-energy term in Eq. (\ref{eq:nm}):
\begin{align}
\mathcal{H}_{\rm s-wave}=\sum_{\bm k} \Delta e^{i\varphi}c^\dag_{{\bm k}\uparrow}c\dag_{-{\bm k}\downarrow} +{\rm c.c.},
\end{align}
where $\varphi$ is the macroscopic phase of the pair potential 
relative to the interface normal of the $p$-wave superconductor.
These three parts are coupled via hopping at the interface.
The magnitude of the hopping at the interface between the NM and the $s$-wave superconductor was chosen to be the same as in the NM.
The SRO-NM interface displays three kinds of hopping:
$t_{xy}$, $t_{yz}$, and $t_{zx}$.
The first, $t_{xy}$, corresponds to the hopping between the NM and the $d_{xy}$ orbital of SRO.
Likewise, $t_{yz}$ ($t_{zx}$) also denotes the interface hopping
between NM and $d_{yz}$ ($d_{zx}$) orbital of SRO.

\par
\begin{figure}[htbp]
	\begin{center}
		\includegraphics[width=8.5cm]{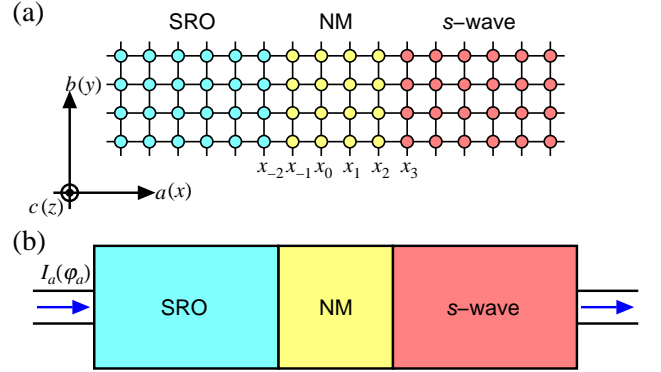}
		\caption{
(a)Lattice model of the junction considered in this paper. 
(b)Schematic illustrations of an SRO (Sr$_{2}$RuO$_{4}$) /NM(normal metal) /$s$-wave superconductor single Josephson junction.
		}\label{sj_fig}
	\end{center}
\end{figure}
We calculated the current-phase relation of the Josephson current in the single junction (see \figref{sj_fig} (a))
based on a lattice Green's function method that takes into account the
Andreev reflection and Andreev bound states at the interface
\cite{Furusaki91,tanaka97}.
For that purpose, we calculated the Green's function in the superconducting SRO/NM/$s$-wave superconductor junction.
These three regions are aligned in the (100) direction, with
the boundaries for the $s$-wave superconductor and SRO located at $x\le x_{-2}$ and $x\ge x_{3}$, respectively.
In the numerical calculations,
four NM layers are inserted between these two superconductors at $x_{-1}\le x\le x_{2}$.
Since we are considering flat interfaces in the ballistic limit,
$k_y$ is a conserved quantity.
In order to obtain the Green's function in this junction,
we first calculated the surface Green's functions of the
semi-infinite SRO and spin-singlet $s$-wave superconductor,
where the surfaces are not coupled to the NM layer. These calculations were based on the recursive Green's function method, using M$\ddot{\rm o}$bius transformation \cite{Umerski}.
Next, we added the two NM layers on these surfaces with the following recursive equation:
\begin{align}
\hat G_{n}^L(k_y,i\omega_l)=(i\omega_l-\hat\varepsilon_n(k_y)-\hat t_{n,n-1}\hat G_{n-1}^L(k_y,i\omega_l)\hat t_{n-1,n})^{-1},\\
\hat G_{n}^R(k_y,i\omega_l)=(i\omega_l-\hat\varepsilon_n(k_y)-\hat t_{n,n+1}\hat G_{n+1}^L(k_y,i\omega_l)\hat t_{n+1,n})^{-1},
\end{align}
where $G_{n}^{LI}(k_y,i\omega_l)$ stands for the surface Green's function for the system on the left (right) side of the interface, with $x\le x_n$ ($x\ge x_n$).
The operators $\hat\varepsilon_n(k_y)$ and $\hat{t}_{n,n-1}$ represent the local and non-local
parts of the Hamiltonian.
Then, we obtained two surface Green's function, defined for $x\le x_0$ and $x\ge x_1$.
These two systems are combined in the equation
\begin{align}
\hat G_{00}(k_y,i\omega_l)=((\hat G_{0}^L(k_y,i\omega_l))^{-1}-\hat t_{01}\hat G_{1}^R(k_y,i\omega_l)\hat t_{10})^{-1},\\
\hat G_{11}(k_y,i\omega_l)=((\hat G_{1}^R(k_y,i\omega_l))^{-1}-\hat t_{10}\hat G_{0}^L(k_y,i\omega_l)\hat t_{01})^{-1}.
\end{align}
Then, we obtained the non-local Green's functions in the $s$-wave/NM/SRO junction as follows:
\begin{align}
	\hat G_{01}(k_y, i\omega_l)&=\hat G_{0}^L(k_y, i\omega_l)\hat t_{01}\hat G_{11}(k_y, i\omega_l)\\
	\hat G_{10}(k_y, i\omega_l)&=\hat G_{1}^R(k_y, i\omega_l)\hat t_{10}\hat G_{00}(k_y, i\omega_l).
\end{align}
The Fourier-transforms of $\hat G_{01}(k_y, i\omega_l)$ and $\hat G_{10}(k_y, i\omega_l)$ are given by
\begin{align}
	\hat G_{01}(k_y,\tau)&=\frac{1}{\beta}\sum_{l}\hat G_{01}(k_y, i\omega_l)e^{-i\omega_l\tau}\\
	\hat G_{10}(k_y,\tau)&=\frac{1}{\beta}\sum_{l}\hat G_{10}(k_y, i\omega_l)e^{-i\omega_l\tau}.
\end{align}
with $\beta=1/(k_{B}T)$ and where $T$ is the temperature. 
The above formulas for $\hat G_{01}(k_y,\tau)$ and $\hat G_{10}(k_y,\tau)$
can be expressed as
\begin{align}
	\hat G_{01}(k_y,\tau)&=-\left< T_{\tau}\left[\hat C_{0}(\tau)\hat C_{1}^{\dagger}\right]\right>\\
	\hat G_{10}(k_y,\tau)&=-\left< T_{\tau}\left[\hat C_{1}(\tau)\hat C_{0}^{\dagger}\right]\right>,
\end{align}
with
\begin{align}
	\hat C_{0}^{\dagger}&=\left(
		\begin{array}{cccc}
			C_{0e\uparrow}^{\dagger}&C_{0e\downarrow}^{\dagger}&C_{0h\uparrow}^{\dagger}&C_{0h\downarrow}^{\dagger}
		\end{array}
	\right)\\
	\hat C_{1}^{\dagger}&=\left(
		\begin{array}{cccc}
			C_{1e\uparrow}^{\dagger}&C_{1e\downarrow}^{\dagger}&C_{1h\uparrow}^{\dagger}&C_{1h\downarrow}^{\dagger}
		\end{array}
	\right).
\end{align}
Thus, we obtained the current-phase relation $I(\varphi)$ by using these $\hat G_{01}(k_y,\tau)$ and $\hat G_{10}(k_y,\tau)$:
\begin{align}
		I(\varphi)=\frac{iet}{\hbar} \int_{-\pi}^{\pi}\mathrm{Tr'}\Bigl[\hat{G}_{01}(k_y,\tau=-0,\varphi)& \nonumber \\
		-\hat{G}_{10}(k_y,\tau=-0,\varphi)&\Bigr] dk_y \nonumber \\
				=\frac{iet}{\hbar} \int_{-\pi}^{\pi}\mathrm{Tr'}\frac{1}{\beta}\sum_{l}\Bigl[\hat G_{01}(k_y, i\omega_l, \varphi)& \nonumber \\
		-\hat G_{10}(k_y, i\omega_l, \varphi)&\Bigr] dk_y,
\end{align}
where $\mathrm{Tr'}$ is a partial sum of the diagonal elements of the Hamiltonian, 
including only those matrix elements that refer to the electron space.
\par
Below, we define the model parameters that were used in the calculations.
For the hopping parameters in SRO, we assumed $t_2/t_{1}=0.395$,
$t_3/t_{1}=1.25$, $t_4/t_{1}=0.125$, and $t_5/t_{1}=0.15$, based on first-principles calculations.
Here, $t_{1}$ is the nearest-neighbor hopping parameter in the $d_{xy}$ orbital in SRO, which first-principles calculations estimate as being approximately 230 meV \cite{Maeno2}.
Furthermore, the chemical potentials in each orbital in SRO, $\mu_{yz}$, $\mu_{zx}$, and $\mu_{xy}$, were chosen to yield the following numbers of electron: $n_{yz}=n_{zx}=n_{xy}=2/3$.
The chemical potential in the normal metal, $\mu_{n}$, was chosen so that the number of electron is $2/3$. 
The magnitude of the spin-orbit interaction in the bulk SRO, expressed as $\lambda$, changes these values.
We set $\lambda=0.3$ for consistency with quasiparticle spectra obtained by angle-resolved photoemission spectroscopy \cite{Maeno2}.
We chose the magnitudes of the pair potential for the $d_{yz}$ and $d_{zx}$ orbitals in SRO to exceed that of the $d_{xy}$ orbital, as determined previously by tunneling spectroscopy \cite{Yada,Kashiwaya11}.
The magnitude of the pair potential in the $d_{yz}$ and $d_{zx}$ orbitals was set to $\Delta_1=0.001t_{1}$.
We set the magnitude of the pair potential for the $d_{xy}$-orbital to $\Delta_2=0.4\Delta_1$.
For the quasi-one-dimensional nature of the pair potential for $d_{xy}$-orbital, we set $\delta=0.1$, based on the ratio of $t_3$ to $t_4$.
\par
We assumed that an $s$-wave superconductor and an NM are described by the same
single-orbital model as that of the $d_{xy}$ orbital in SRO.
We set their chemical potentials $\mu_n$ to the same level as the $d_{xy}$ orbital in SRO, in the absence of spin-orbit interaction in the bulk SRO.
The magnitude of the pair potential of the $s$-wave superconductor was set to $\Delta_s=10\Delta_1$.
The magnitude of the Rashba spin-orbit interaction at the
interface between NM and SRO, $\lambda_R$, depends on the
microscopic electronic properties of the junction and was set to 
$0.3$ in this study.
%

\section{Results}\label{sec3}
\subsection{current phase relation}
\begin{figure}[htbp]
\begin{center}
\includegraphics[width=8.5cm]{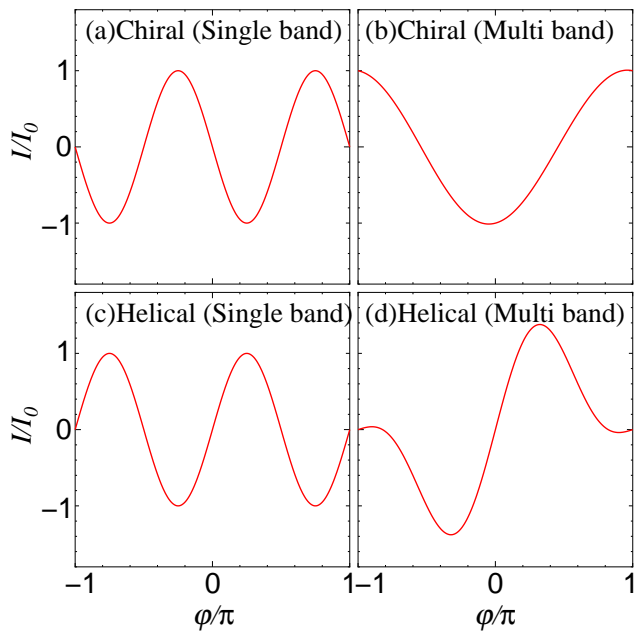}
\caption{
Current-phase relation in the absence of interface Rashba spin-orbit interaction ($\lambda_{R}$) for
(a) the chiral $p$-wave ($E_u$) in the single-band model,
(b) the chiral $p$-wave ($E_u$) in the multi-band model,
(c) the helical $p$-wave ($A_{1u}$) in the single-band model,
and (d) the helical $p$-wave ($A_{1u}$) in the multi-band model
}\label{cpr_wo_gra}
\end{center}
\end{figure}

Figure \ref{cpr_wo_gra} shows the current-phase relation in the absence of interface Rashba spin-orbit interaction.
Here, the Josephson current $I(\varphi)$ is decomposed into the 
Fourier series
\begin{eqnarray}
	I(\varphi)=\sum_{n=1}^{\infty}{I_n^s\sin(n\varphi)+I_n^c\cos(n\varphi)}.\label{Fourier}
\end{eqnarray}
It is then normalized by $I_0$, the maximum value of the Fourier coefficients.
Table \ref{cpr_wo_tab} shows which of the Fourier coefficients have nonzero values.
%
\begin{table}[htbp]
\begin{center}
\begin{tabular}{|l|c|c|c|c|} \hline
 &\ \ $I_1^s$\ \  & \ \ $I_1^c$\ \  & \ \ $I_2^s$\ \  & \ \ $I_2^c$\ \  \\ \hline
(a)Chiral(single-band) & $-$ & $-$ & $\surd$ & $-$\\
(b)Chiral(multi-band) & $\surd$ & $\surd$ & $\surd$ & $\surd$\\
(c)Helical(single-band) & $-$ & $-$ & $\surd$ & $-$ \\
(d)Helical(multi-band) & $\surd$ & $-$ & $\surd$ & $-$ \\ \hline
\end{tabular}
\caption{Fourier series of current-phase relation in the absence of interface Rashba spin-orbit interaction. $\surd$ ($-$) denotes
coefficients with a nonzero (zero) value.}
\label{cpr_wo_tab}
\end{center}
\end{table}
\par
As shown in Figs. \ref{cpr_wo_gra}(a) and (c), the Josephson current is almost proportional to $\sin(2\varphi)$ in the case where the first-order Josephson coupling is absent.
In fact, Table \ref{cpr_wo_tab} shows that only the sinusoidal terms with an even-number order are nonzero.
On the other hand, odd-order terms are nonzero in the case of the multi-band model, as shown in Fig. \ref{cpr_wo_gra} and Table \ref{cpr_wo_tab} (b) and (d).
We confirmed that these odd-order terms are zero in the absence of spin-orbit interaction (LS coupling) in bulk SRO.
We note that the cosine terms appear in the chiral $p$-wave case but are absent in the helical $p$-wave case.
The cosine terms in the chiral $p$-wave case are nonzero even in the absence of Rashba spin-orbit coupling $\lambda_{R}$.
This is because the hopping integral $t_5$ ($i.e.$, corresponding to
inter-orbital hopping between the $d_yz$ and $d_xz$ orbitals) 
is nonzero and spin-orbit coupling in bulk SRO $\lambda$ enhances the magnitude of the cosine terms.
When the opposite chirality of the pair potential is chosen with 
\begin{align}
&\begin{cases}
\bm d_{Eu}^{yz}=\hat z\Delta_1(\delta\sin{k_x}-i\sin{k_y}),\\
\bm d_{Eu}^{zx}=\hat z\Delta_1(\sin{k_x}-i\delta\sin{k_y}),\\
\bm d_{Eu}^{xy}=\hat z\Delta_2(\sin{k_x}-i\sin{k_y}), \\
\end{cases}
\end{align}
the signs of $I_{1}^{c}$ and $I_{2}^{c}$ are reversed. 
\par
\begin{figure}[htbp]
\begin{center}
\includegraphics[width=8.5cm]{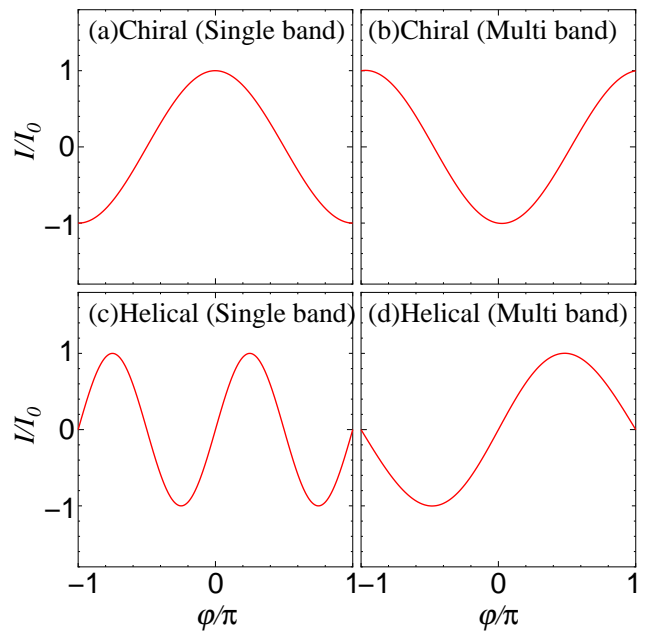}
\caption{
Current-phase relation $I(\varphi)$
in the presence of interface Rashba spin-orbit interaction ($\lambda_{R}>0$) for
(a) the chiral $p$-wave ($E_u$) in the single-band model,
(b) the chiral $p$-wave ($E_u$) in the multi-band model,
(c) the helical $p$-wave ($A_{1u}$) in the single-band model,
and (d) the helical $p$-wave ($A_{1u}$) in the multi-band model.
}\label{cpr_w_gra}
\end{center}
\end{figure}
\begin{table}[htbp]
\begin{center}
\begin{tabular}{|l|c|c|c|c|} \hline
 &\ \ $I_1^s$\ \  & \ \ $I_1^c$\ \  & \ \ $I_2^s$\ \  & \ \ $I_2^c$\ \  \\ \hline
(a) Chiral(single-band) & $-$ & $\surd$ & $\surd$ & $-$ \\
(b) Chiral(multi-band) & $\surd$ & $\surd$ & $\surd$ & $\surd$\\
(c) Helical(single-band) & $-$ & $-$ & $\surd$ & $-$ \\
(d) Helical(multi-band) & $\surd$ & $-$ & $\surd$ & $-$ \\ \hline
\end{tabular}
\caption{
Fourier series of the current-phase relation in the presence of interface Rashba spin-orbit interaction
}
\label{cpr_w_tab}
\end{center}
\end{table}\noindent
We plot the current-phase relations in the presence of interface Rashba spin-orbit coupling ($\lambda_{R}>0$) in Fig. \ref{cpr_w_gra}.
Figure \ref{cpr_w_gra} (c) shows no qualitative difference between the Josephson currents in the presence or absence of interface Rashba spin-orbit interaction, in the
single-band model and in the case of helical pairing.
On the other hand, cosine terms appear as a result of the interface Rashba spin-orbit coupling in the case of the chiral $p$-wave shown in Fig. \ref{cpr_w_gra} (a) 
\cite{asano03}.
By contrast, there is no qualitative difference between the current-phase relations in the presence or absence of interface Rashba spin-orbit interaction in the multi-band model, as shown in Figs. \ref{cpr_w_gra} (b,d) and Table \ref{cpr_w_tab} (b,d).
In the most general case, where both the interface Rashba spin-orbit interaction and bulk LS coupling in the multi-band model exist,
we observe a qualitative difference between the chiral and helical 
$p$-wave cases. The cosine terms $I_{1}^{c}$ and $I_{2}^{c}$
appear only in the case of chiral $p$-wave pairing.
This difference is due to the broken TRS that occurs in chiral 
$p$-wave pairing. 
In the following calculations for various junctions, we considered the interface Rashba spin-orbit interactions and used the multi-band model.
%
\begin{figure}[htbp]
\begin{center}
\includegraphics[width=8.0cm]{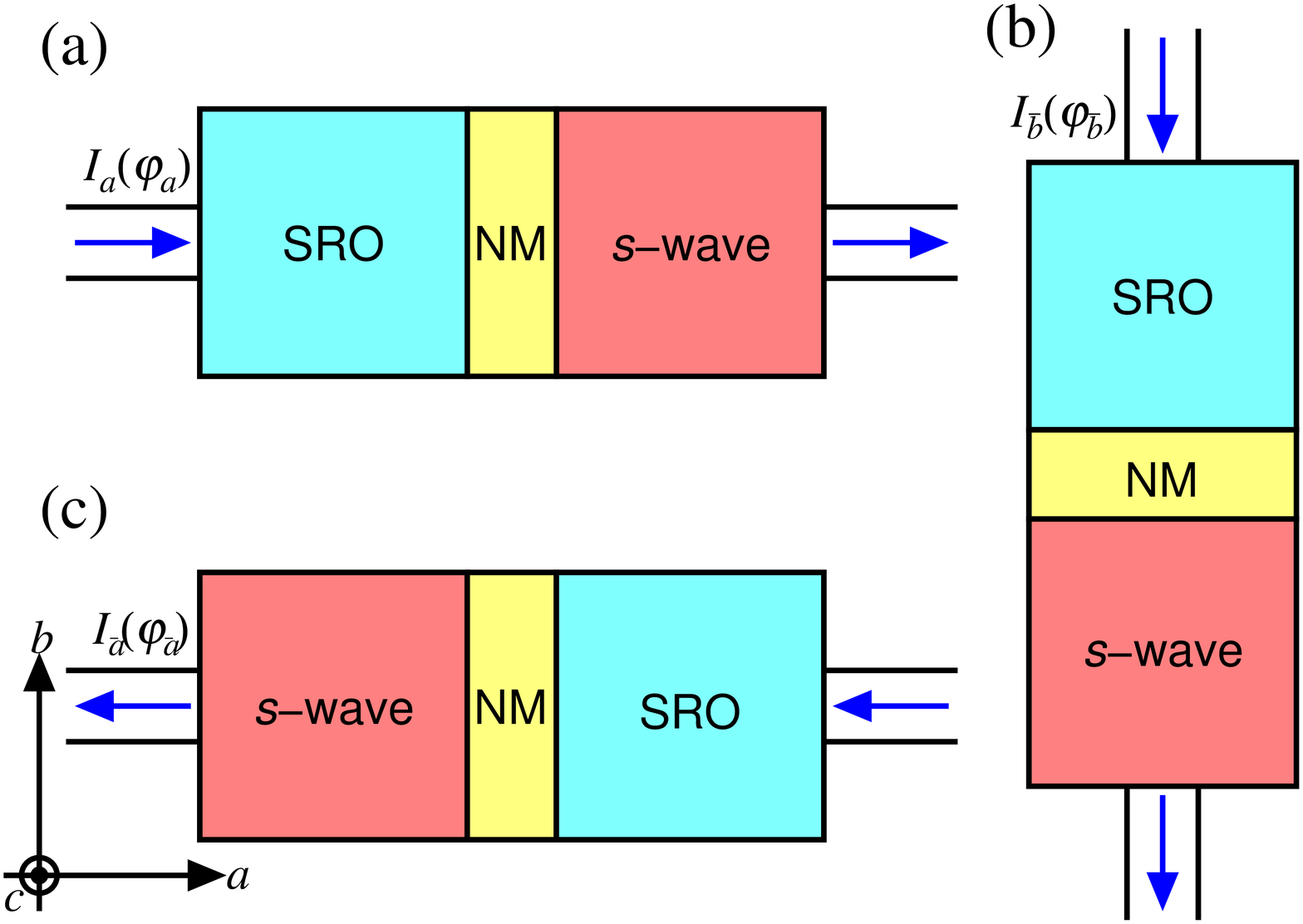}
\caption{
Schematic illustrations of the SRO /NM/$s$-wave superconductor single Josephson junctions considered 
in this paper. Current-phase relations in junctions (a)-(c) were calculated independently. The results were then combined to calculate the magnetic-field dependence of the corner junction, corner SQUID, and symmetric SQUID.  }
\label{sjs_fig}
	\end{center}
\end{figure}
\begin{table}[htbp]
\begin{center}
\begin{tabular}{|l|c|} \hline
Type of pairing & Relation between $I_a(\varphi_a)$ and $I_{\overline b}(\varphi_{\overline b})$ \\ \hline
Chiral($E_u$) & $I_a(\varphi_{a})=-I_{\overline b}(-\varphi_{\overline b}+\pi/2)$ \\ \hline 
Helical($A_{1u}$, $B_{2u}$) & $I_a(\varphi_{a})=I_{\overline b}(\varphi_{\overline b})$ \\ \hline
Helical($A_{2u}$, $B_{1u}$) & $I_a(\varphi_{a})=I_{\overline b}(\varphi_{\overline b}+\pi)$ \\ \hline
\end{tabular}
\caption{
Relations between $I_a(\varphi_a)$ and $I_{\overline b}(\varphi_{\overline b})$ shown in Fig. \ref{sj_fig} for chiral($E_u$), helical ($A_{1u}$, $B_{2u}$), and helical ($A_{2u}$, $B_{1u}$) pairings.
}
\label{cpabc_tab}
\end{center}
\end{table}\noindent
\par
In order to take into account the corner structure of the junction, we show the relation between the current phase relations in different orientations in Table III.
The orientation dependence affects the maximum Josephson current in a corner junction or SQUID when it is written as a function of the external magnetic flux $\Phi$.
Although the calculation of the $\Phi$ dependence will be shown in next subsection, 
we first show the relation between $I_a(\varphi_a)$ and $I_{\overline b}(\varphi_{\overline b})$ indicated in
Fig. \ref{sjs_fig}.
This relation depends on the pairing symmetries specified in \tabref{cpabc_tab}
This relation in chiral $p$-wave pairing is different from that in helical $p$-wave pairing.
Furthermore, in the helical $p$-wave cases, the relation between $I_a(\varphi_a)$ and $I_{\overline b}(\varphi_{\overline b})$ depends 
on the irreducible representations of the pair potentials.
This affects the properties of the corner junction or corner SQUID, as shown in the next subsection.
Next, we show the relation between the $I_a(\varphi_a)$ and $I_{\overline a}(\varphi_{\overline a})$ indicated in Fig. \ref{sj_fig}.
The equation $I_{a}(\varphi)=I_{\overline a}(\varphi+\pi)$ is valid for all pairings.
This fact influences the properties of a symmetric SQUID.

\subsection{Magnetic-field dependence of the maximum Josephson current 
in various junctions}
In this subsection, we calculate the magnetic-field dependence of the maximum 
Josephson current in corner junctions, corner SQUIDs, and 
symmetric SQUIDs. 
\begin{figure}[htbp]
	\begin{center}
		\includegraphics[width=6cm]{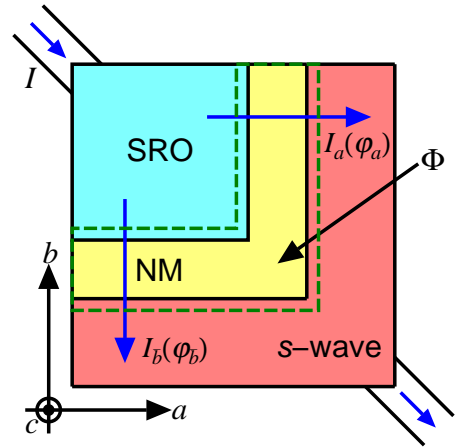}
		\caption{
		Schematic illustration of an SRO/NM/$s$-wave corner junction
		}\label{cj_fig}
	\end{center}
\end{figure}
We calculated the relation between the external magnetic flux $\Phi$ and the
maximum Josephson current $I_{c}$ by a standard method.
In the Josephson junctions shown in Figs. \ref{sj_fig} and \ref{sjs_fig},
we assumed that the external magnetic field was applied parallel to the $z$-axis.
The vector potential is then given by
\begin{eqnarray}
	\bm{A}=A_y(x)\bm{y}.
\label{vector_equ}
\end{eqnarray}
On the other hand, the phase $\gamma$ of the pair potential obeys
\begin{eqnarray}
	\nabla\gamma=\frac{m^*\bm v_s}{\hbar}+\frac{2\pi}{\Phi_0}\bm{A}.
\label{gamma_equ}
\end{eqnarray}
Since the magnetic field is screened inside the
superconductor because of the
Meissner effect, $A_y(x)$ takes the constant value $A_y(\infty)$ found at locations far from the
interface.
Using these properties, we integrated both sides of the $y$ component of \equref{gamma_equ} with respect to $y$.
\begin{eqnarray}
	\gamma(y)=\gamma(0)+\frac{2\pi}{\Phi_0}A(\infty)y
\end{eqnarray}
The phase difference between the $s$-wave superconductor and the SRO is therefore given by
\begin{eqnarray}
	\varphi(y)=\varphi(0)+\frac{2\pi}{\Phi_0}\left[A_2(\infty)-A_1(\infty)\right]y.
\end{eqnarray}
Here, $A_{1}$ and $A_{2}$ represent the vector potentials far from the interface in 
the SRO and $s$-wave superconductor, respectively. 
The Fourier components of the Josephson current, 
$I_{n}^{s}$ and $I_{n}^{c}$, defined in eq. (\ref{Fourier}),
were obtained in the previous subsection in the absence of a 
magnetic field.
%
In the presence of a magnetic field, the Josephson current becomes a function of 
$y$. We integrated this function with respect to $y$:
\begin{eqnarray}
\label{iphi1_equ}
	&I&(\Phi, \varphi(0))=Z\int_{-Y/2}^{Y/2}I(y)dy\\ \notag \nonumber
	&=&YZ\sum_{n=1}^{\infty}\Biggl\{\frac{\sin(n\pi\Phi/\Phi_0)}{n\pi\Phi/\Phi_0}\left[I_{n}^{s}\sin(n\varphi(0))+I_{n}^{c}\cos(n\varphi(0))\right]\Biggr\},\end{eqnarray}
where, $Y$ and $Z$ are the sizes of the junction.
It is evident that 
Eq.(\ref{iphi1_equ}) 
displays a periodicity of 2$\pi$ with respect to $\varphi(0)$.
Therefore, by changing $\varphi(0)$ over the range $-\pi\leqq\varphi(0)\leqq\pi$, the maximum Josephson current $I_{c}$ can be obtained as a function of the external magnetic flux $\Phi$.
\par
Next, we calculated the maximum Josephson current $I_{c}$ in the corner junction shown in Fig. \ref{cj_fig} as a function of $\Phi$, using a similar approach to that 
described in [\onlinecite{Harlingen}].
We obtained the current-phase relations $I_a(\varphi_a)$ and $I_{\overline b}(\varphi_{\overline b})$ indicated in Fig. \ref{cj_fig}.
By calculating the following equation instead of 
Eq. (\ref{iphi1_equ}), we obtained 
the maximum Josephson current $I_{c}$ as a function of the external 
magnetic flux $\Phi$ based on $I(\Phi, \varphi(0))$, given by
\begin{eqnarray}
	I(\Phi, \varphi(0))=Z\left[\int_{0}^{Y/2}{I_a(y)dy}+\int_{-Y/2}^{0}{I_{\overline b}(y)dy}\right].
\nonumber
\end{eqnarray}
\par
\begin{figure}[htbp]
	\begin{center}
		\includegraphics[width=8cm]{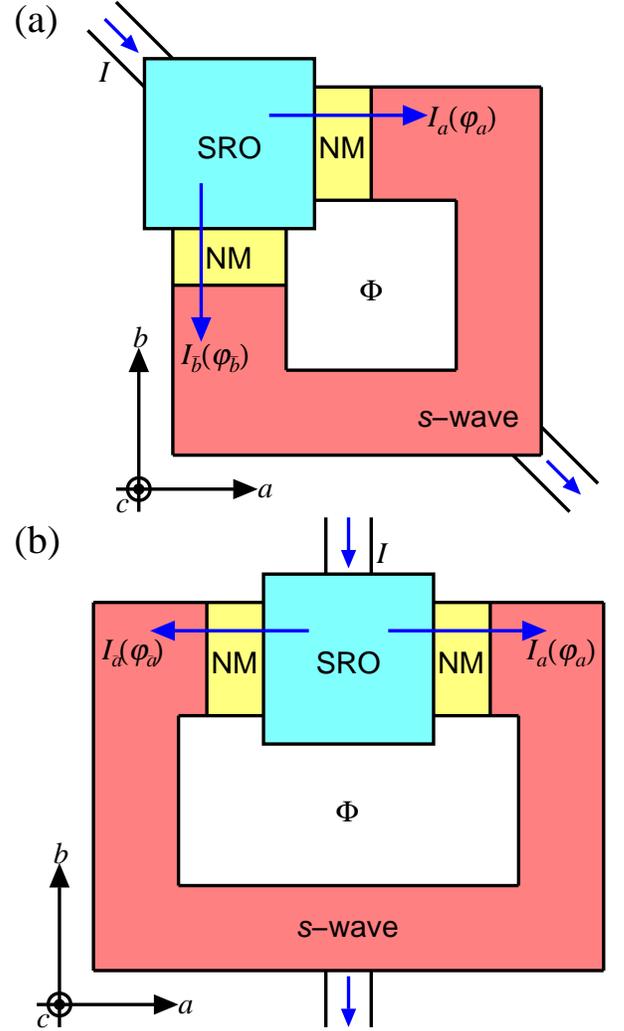}
		\caption{
		Schematic illustration of SRO/NM/$s$-wave SQUIDS:
                (a) corner SQUID and (b) symmetric SQUID.
		}\label{squid_fig}
	\end{center}
\end{figure}
Finally, we calculated the maximum Josephson current 
$I_{c}$ as a function of the external magnetic flux $\Phi$ in the two types of 
SQUID  
shown in Fig. \ref{squid_fig}.
The macroscopic phase differences of the two superconductors $\varphi_a$ and $\varphi_{\overline b}$
obey the following relation:
\begin{eqnarray}
	\varphi_{\overline b}-\varphi_a=\frac{2\pi\Phi}{\Phi_0}.
\end{eqnarray}
The total current in these parallel circuits is therefore given by
\begin{eqnarray}
	I(\Phi, \varphi)=I_a(\varphi)+I_{\overline b}(\varphi+\frac{2\pi\Phi}{\Phi_0}).
	\label{iphi_equ}
\end{eqnarray}
By evaluating the maximum value of 
\equref{iphi_equ} 
for a given external magnetic flux $\Phi$, we obtained the maximum Josephson current as a function of $\Phi$.
\par
%
\begin{figure}[htbp]
\begin{center}
\includegraphics[width=8.5cm]{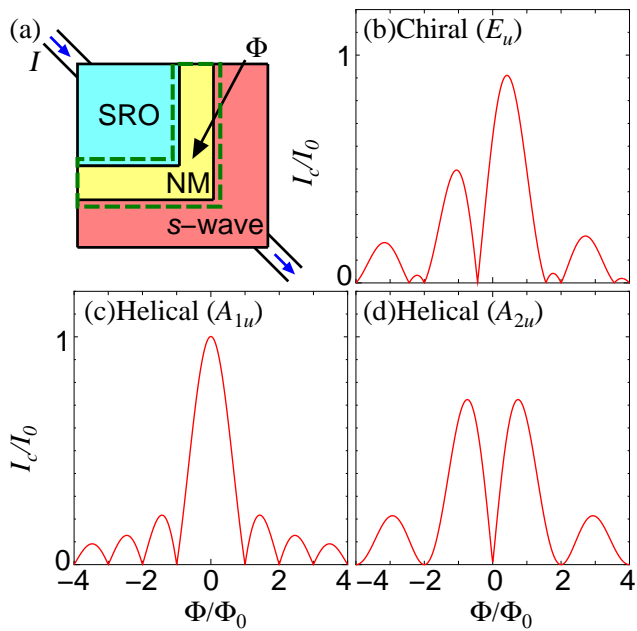}
\caption{
Fraunhofer pattern in the SRO/NM/$s$-wave.
(a) Schematic illustration of a corner junction, and the corresponding
Fraunhofer pattern for 
(b) chiral ($E_u$) pairing, 
(c) helical ($A_{1u}$) pairing,
 and
(d) helical ($A_{2u}$) pairing.
}\label{cj_gra}
\end{center}
\end{figure}
\begin{table}[htbp]
\begin{center}
\begin{tabular}{|l|c|c|} \hline
Type of pairing & $\Phi$ dependence & zero points of $I(\Phi)$ \\ \hline
(a)Chiral($E_u$) & asymmetric & $\pm2\Phi_0,\,\pm4\Phi_0,\,\cdots$ \\ \hline
(b)Helical($A_{1u}$, $B_{2u}$) & symmetric & $\pm\Phi_0,\,\pm2\Phi_0,\,\cdots$ \\ \hline
(c)Helical($A_{2u}$, $B_{1u}$) & symmetric & $\pm2\Phi_0,\,\pm4\Phi_0,\,\cdots$ \\ \hline
\end{tabular}
\caption{
$\Phi$ dependence and zero points of $I(\Phi)$ in an SRO/NM/$s$-wave
corner junction for (b) chiral($E_u$), (c) helical($A_{1u}$, $B_{2u}$), and (d) helical($A_{2u}$, $B_{1u}$) pairings in SRO.
Schematic illustration of corner junction (a). }
\label{cj_tab}
\end{center}
\end{table}\noindent
\par
The $I_{c}$ functions 
for the corner junction of SRO are plotted in Fig. \ref{cj_gra}.
In the cases of the helical $p$-wave, the positions of the minima depend on the $d$-vector as shown in Figs. \ref{cj_fig}(b) and (c).
This is because the relation between the Josephson currents $I_a(\varphi_a)$ and $I_{\overline b}(\varphi_{\overline b})$ in Fig. \ref{cj_gra} is different for each pairing symmetry.
In particular, $I_a(\varphi) = I_{\overline b}(\varphi)$ for the $A_{1u}$ and $B_{2u}$ pairings, while $I_a(\varphi)=I_{\overline b}(\varphi+\pi)$ for the $A_{2u}$ and $B_{1u}$ pairings.
For all the helical $p$-wave cases, the Fraunhofer patterns are symmetric functions of $\Phi$.
On the other hand, $I(\Phi)$ is not a symmetric function of $\Phi$ for chiral $p$-wave pairing.
This difference results from the existence of the cosine terms in the current-phase relation.
In other words, the broken TRS causes the
asymmetry of $I_{c}=I_{c}(\Phi)$, $i.e.$, $I_{c}(\Phi) \neq I_{c}(-\Phi)$.
These results are summarized in Table \ref{cj_tab}.
As seen from this table, there are qualitative differences between the helical and chiral $p$-wave pairings.
The asymmetry of the Josephson current is due to the existence of cosine terms in the current-phase relation for the chiral $p$-wave pairings.
These cosine terms can be nonzero unless both $\lambda$ and $\lambda_{R}$ are nonzero
owing to the presence of inter-orbital hopping in the multi-band model.
The magnitudes of these cosine terms and the resulting asymmetry of $I(\Phi)$
are enhanced by the spin-orbit interactions, expressed through $\lambda$ and $\lambda_{R}$.
\par
%
\begin{figure}[htbp]
\begin{center}
\includegraphics[width=8.5cm]{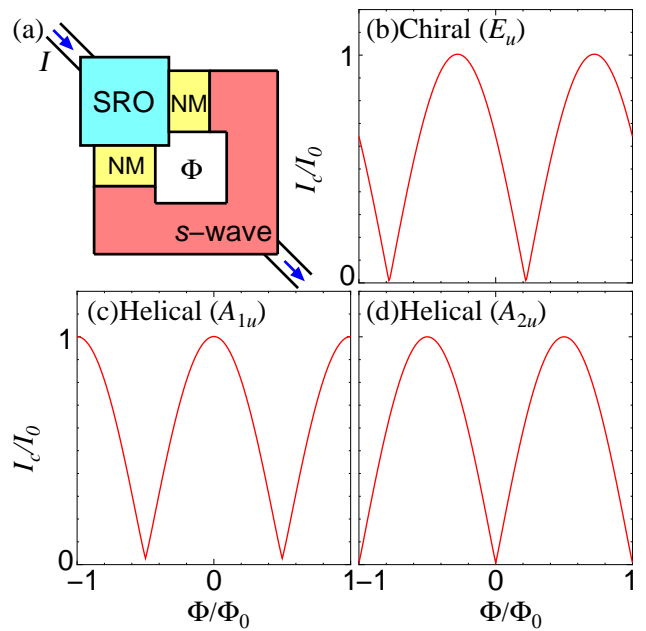}
\caption{
(a) Maximum Josephson current $I_{c}$ in a corner SQUID for (b) chiral ($E_u$), (b)helical ($A_{1u}$), and (c) helical ($A_{2u}$) pairings.
}\label{cs_gra}
\end{center}
\end{figure}
\begin{table}[htbp]
\begin{center}
\begin{tabular}{|l|c|c|} \hline
Type of pairing & $\Phi$ dependence & Period \\ \hline
(a) Chiral & asymmetric & $\Phi_0$ \\ \hline
(b, c) Helical & symmetric & $\Phi_0$ \\ \hline
\end{tabular}
\caption{
$\Phi$ dependence and period of the maximum Josephson current
$I_{c}$ in a corner SQUID
}
\label{cs_tab}
\end{center}
\end{table}\noindent

Next, we discuss $I_{c}$ in the corner SQUID shown in
Fig. \ref{cs_gra}.
This $I_{c}$ is
symmetric or asymmetric with respect to $\Phi$
for the helical and chiral cases, respectively.
As in the case of the SRO/NM/$s$-wave corner junction,
the existence of the cosine terms in the current-phase
relation in chiral pairing causes the asymmetry of $I_{c}(\Phi)$.
The chiral pairing is consistent with a 
previous study based on a single-band model \cite{Asano2005}. 
In the cases of helical pairing,
the position of the maximum or minimum in $I_{c}(\Phi)$
depends on the pairing symmetry (irreducible representation), $i.e.$,
the $d$-vector as shown in Figs. \ref{cs_gra}(b) and (c).
We note that the $\Phi_{0}$ periodicity in the 
helical pairing case appears only for a three-band model. 
\par
%
\begin{figure}[htbp]
\begin{center}
\includegraphics[width=8.5cm]{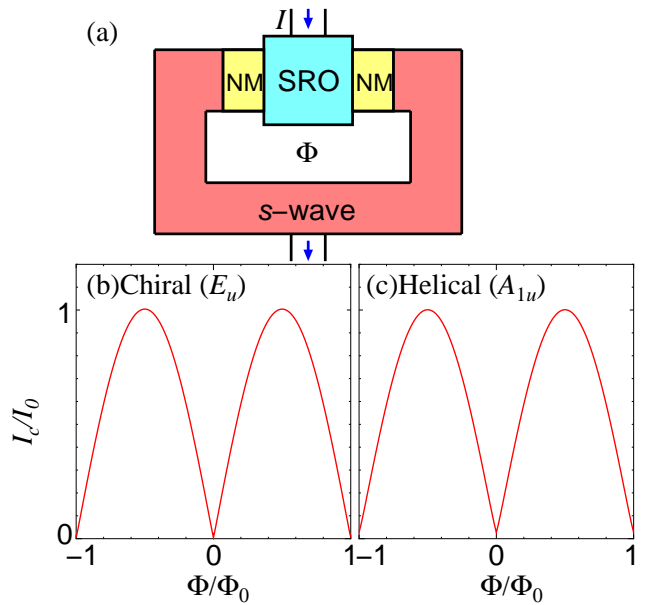}
\caption{
(a) Symmetric SQUID and the corresponding $I_{c}$ for (b) chiral ($E_u$) and (c) helical ($A_{1u}$) pairings
}
\label{ss_gra}
\end{center}
\end{figure}
\begin{table}[htbp]
\begin{center}
\begin{tabular}{|l|c|c|} \hline
Type of pairing & $\Phi$ dependence & Period \\ \hline
(a) Chiral & symmetric & $\Phi_0$ \\ \hline
(b) Helical & symmetric & $\Phi_0$ \\ \hline
\end{tabular}
\caption{
$\Phi$ dependence and period of the maximum Josephson current $I_{c}$
in a symmetric SQUID.
}
\label{ss_tab}
\end{center}
\end{table}
Finally, we consider the case of the so-called symmetric SQUID \cite{Geshkenbein}. 
Figure \ref{ss_gra} shows the $\Phi$ dependence of
$I_{c}$ in the symmetric SQUID shown in Fig. \ref{squid_fig}(b).
In this junction, there is no qualitative difference between the cases of chiral and helical pairing since
$I_a(\varphi)=I_{\overline b}(\varphi+\pi)$ is satisfied.
The resulting Josephson current $I_{c}$ is symmetric for both chiral and helical pairings, including in the presence of the cosine terms.
Thus, we do not find any qualitative difference in $I_{c}$ for the chiral and helical pairings in this symmetric SQUID.
%
\section{Discussion and Summary}\label{sec4}
\begin{figure}[htbp]
\begin{center}
\includegraphics[width=8.5cm]{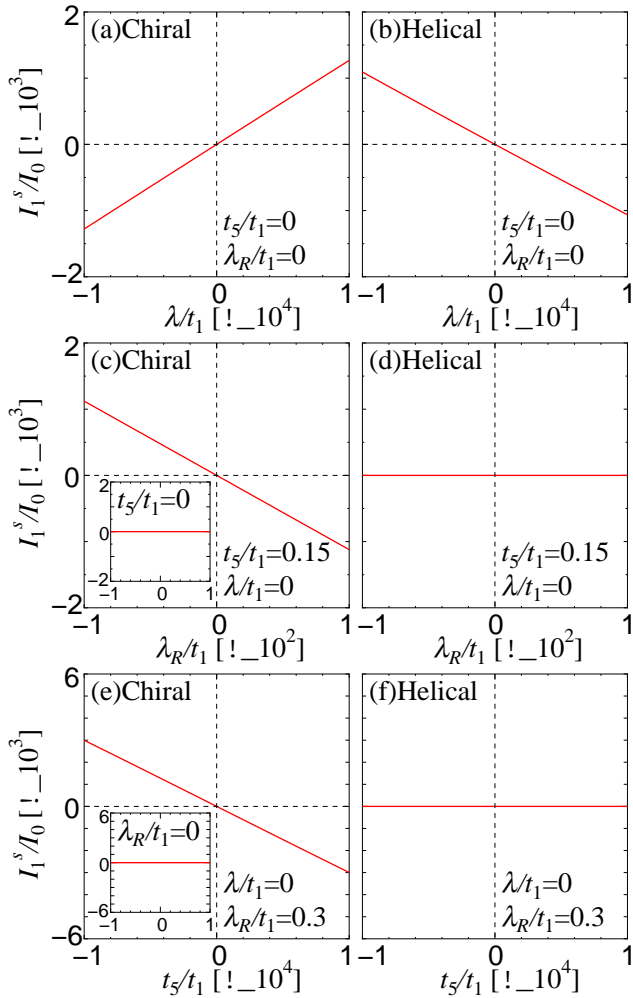}
\caption{$I_1^s$ (the coefficient of $\sin(\varphi)$ in the Fourier series of the current-phase relation in the junction) is plotted as a function of
$\lambda$ (a, b), $\lambda_{R}$ (c, d), and $t_{5}$ (e, f).
Chiral pairing applies in (a), (c), and (e), and
helical pairing with A$_{1u}$ symmetry in (b), (d), and (f).
$t_5=0$ and $\lambda_R=0$ in (a) and (b).
$t_5/t_1=0.15$ and $\lambda=0$ in (c) and (d).
$\lambda=0$ and $\lambda/t_1=0.3$ in (e) and (f).
}
\label{relation_gra}
\end{center}
\end{figure}
Here, we discuss the multi-band effect on the Josephson current in the present calculations, starting with the chiral $p$-wave case.
As shown in TABLES \ref{cpr_wo_tab} and \ref{cpr_w_tab},
the spin-orbit interaction in the bulk SRO ($\lambda$) and the interface Rashba spin-orbit interaction ($\lambda_R$)
generate $I_1^s$ for chiral $p$-wave pairing.
We found that the coefficient of the $\sin(\phi)$ term $I_1^s$ has the form
\begin{equation}
I_1^s=\alpha \lambda+\beta t_5 \lambda_R +O(\lambda^2)+O(t_5^2 \lambda_R^2),\label{eq:i1s}
\end{equation}
which is confirmed by Fig. \ref{relation_gra}.
This form suggests that $\lambda$ directly induces $I_1^s$, whereas
the existence of inter-orbital hopping $t_5$ is needed to produce $I_1^s$ from $\lambda_R$.
In the single-band model, $I_1^s$ is absent while $I_1^c$ is induced by $\lambda_R$ in chiral $p$-wave pairing.
In the multi-orbital model, $t_5$ induces the effective phase shift of the pair potential.
A part of $I_1^c$ is then converted to $I_1^s$ by $t_5$.
Thus, we can conclude that the existence of $I_1^s$ results from the multi-band model in SRO.
This term becomes dominant
in the limit of low transmissivities,
where the higher-order Josephson couplings are strongly suppressed. \par
Next, we discuss the helical $p$-wave case, where 
$I_1^s$ is given by
\begin{equation}
I_1^s=\alpha \lambda + O(\lambda^2). 
\label{eq:i1sh}
\end{equation}
This is because $I_1^c$ does not exist in the single-band model owing to the TRS of the helical $p$-wave pairing.
Since $t_5$ only gives the effective phase shift of the pair potential, $I_1^s$ cannot be produced by $\lambda_R$.
On the other hand, $\lambda$ directly induces $I_1^s$ in a similar manner as in the case of the chiral $p$-wave pairing. \par
In summary, we have studied Josephson currents in SRO/NM/$s$-wave junctions.
We found that the first-order Josephson coupling is induced by the spin-orbit interaction
for the cases of both chiral and helical $p$-wave pairings.
Note that the $\sin(\varphi)$ term, which is absent in the
single-band model, appears as a result of the spin-orbit interaction
and inter-band hopping.
In the case of helical pairing, the first-order Josephson term appears 
only in the three-band model. 
Owing to the existence of the first-order Josephson coupling,
the period of the Josephson current, as 
the magnetic flux $\Phi$ is varied, is expected to become the period of the conventional junctions.
For the case of chiral $p$-wave pairing, the Josephson current shows asymmetric behavior in the corner junction and the corner SQUID,
owing to broken TRS.
This asymmetry is enhanced by the spin-orbit interaction in the bulk SRO or at the interface in the junction.
Since the magnitude of the spin-orbit interaction in SRO is not very small, it is possible to detect the asymmetry experimentally if the TRS breaking by chiral pairing is realized. \par
In this paper, we assumed ballistic junctions with flat interfaces.
Surface roughness and impurity scattering are known to influence 
charge transport in spin-triplet
$p$-wave superconductor junctions \cite{Bakurskiy,LuBo}.
In particular, the odd-frequency spin-triplet $s$-wave component
generated near the interface induces an anomalous proximity effect \cite{Proximityp,Proximityp2}, and the
resulting Josephson current displays a low-temperature anomaly \cite{Proximityp,Proximityp2,Proximityp3}.
Taking into account the impurity-scattering effect in the multi-band
model is an interesting prospect for future work.

%
%
\begin{acknowledgments}
This work was supported
by a Grant-in-Aid for Scientific Research on Innovative
Areas, Topological Material Science (Grants No. JP15H05851,
No. JP15H05852, and No. JP15H05853), a Grant-in-Aid for
Scientific Research B (Grant No. JP15H03686), a Grant-in-Aid
for Challenging Exploratory Research (Grant No. JP15K13498)
from the Ministry of Education, Culture, Sports, Science, and
Technology, Japan (MEXT); Japan-RFBR JSPS Bilateral Joint
Research Projects/Seminars (Grants No. 15-52-50054 and No.
15668956); Dutch FOM; the Ministry of Education and
Science of the Russian Federation, Grant No. 14.Y26.31.0007;
and by the Russian Science Foundation, Grant No. 15-
12-30030.
\end{acknowledgments}


\appendix

\bibliography{josephson}

\end{document}